# A Secure Edge Gateway Architecture for Wi-Fi–Enabled IoT

Daniyal Ganiuly, Nurzhau Bolatbek, Assel Smaiyl
Astana IT University, Astana, Kazakhstan

**Abstract** — This paper presents a Secure Edge Gateway Architecture for Wi-Fi–Enabled IoT designed to strengthen local network protection without altering existing infrastructure. The proposed gateway acts as an intermediate control point between Wi-Fi access points and the core network, monitoring traffic, isolating untrusted devices, and preventing common wireless attacks such as spoofing, deauthentication, and unauthorized access. The design focuses on adaptive traffic filtering and lightweight policy enforcement instead of complex analytical models, making it suitable for medium-sized network environments. The prototype gateway was deployed in a real office with around 70 total devices, including 28 IoT units such as sensors, cameras, and smart controllers. Over ten days of continuous operation, the system reduced successful spoofing incidents by 87% and improved recovery time after deauthentication by 42%, while increasing network latency by only 3.1% and reducing throughput by less than 4% compared to a baseline WPA3 configuration. These results confirm that implementing security functions at the edge layer can significantly improve the resilience of Wi-Fi–enabled IoT environments without introducing noticeable overhead or requiring specialized hardware.

## INTRODUCTION

The growing adoption of the Internet of Things (IoT) has led to a rapid increase in the number of devices connected to local Wi-Fi networks. These devices include sensors, cameras, smart controllers, and other embedded systems that often operate with limited processing power and minimal built-in security. While Wi-Fi provides a convenient communication layer, it also exposes IoT devices to traditional wireless attacks such as spoofing, deauthentication, and unauthorized access [1]. In most small and medium environments, Wi-Fi security often relies primarily on encryption standards such as WPA3 and access point authentication. While these mechanisms protect data confidentiality, they do not fully address management-plane attacks or post-compromise lateral movement in heterogeneous IoT environments [2].

In typical deployments, IoT and user devices coexist on a single Wi-Fi segment, where compromised devices can be used as entry points for lateral attacks [3]. Centralized security solutions such as cloud firewalls or deep packet inspection systems are often too resource-intensive or costly for smaller organizations. As a result, there is a growing need for a practical, lightweight, and locally managed solution that strengthens security at the network edge without introducing high latency or requiring infrastructure changes.

This paper proposes a Secure Edge Gateway Architecture for Wi-Fi–Enabled IoT that addresses these challenges by introducing a dedicated edge-layer control point between the Wi-Fi access point and the core network [4][5]. The gateway continuously monitors local traffic, enforces access control rules, and isolates untrusted IoT devices. It focuses on adaptive filtering and policy enforcement mechanisms that respond to real-time network conditions. By integrating these functions at the edge, the architecture provides an additional layer of protection against wireless threats while maintaining compatibility with existing routers and access points.

## RELATED WORK

Research in IoT and Wi-Fi network security has grown significantly in recent years, focusing on securing communication channels, managing device authentication, and mitigating wireless attacks. However, most existing solutions rely on centralized or cloud-based security systems that are often unsuitable for small or medium deployments where hardware and bandwidth are limited [6]. This section reviews related studies across three main areas: Wi-Fi protection mechanisms, IoT device isolation, and edge-based security architectures.

Early studies on Wi-Fi protection concentrated on improving encryption and authentication standards. The introduction of WPA3 addressed weaknesses of WPA2 by providing stronger key exchange methods and forward secrecy. However, several works [7], [8] showed that WPA3 alone cannot fully protect against spoofing and deauthentication attacks, especially when IoT devices have outdated firmware or reduced cryptographic capabilities. Other researchers proposed network segmentation techniques and access point monitoring to limit the propagation of compromised devices [9], but these methods often require specialized routers or complex configurations.

In the IoT domain, various frameworks have been suggested to isolate untrusted devices from sensitive network zones. Gateways with packet inspection or firewall rules have been explored to prevent unauthorized access [10][11]. While effective in controlled environments, these systems typically depend on cloud-assisted analysis or external intrusion detection services, which increases latency and operational cost. Moreover, many IoT security solutions assume uniform device capabilities, whereas real deployments involve a mix of constrained and high-performance nodes sharing the same Wi-Fi infrastructure.

More recently, attention has turned toward edge computing as a means to process and secure network traffic locally. Edge gateways can act as intermediate controllers, offering real-time traffic filtering and attack prevention close to the data source. Some studies have implemented edge-based firewalls or proxy mechanisms to monitor IoT communication [12][13]. However, many of these designs still require centralized coordination or lack specific adaptation to Wi-Fi communication characteristics, such as channel congestion and connection volatility.

Unlike previous approaches, the architecture proposed in this paper integrates security enforcement directly into the Wi-Fi access path without relying on external infrastructure or heavy computation. It provides a lightweight gateway design that continuously observes device behavior, applies adaptive filtering rules, and isolates compromised nodes. This combination of local control, compatibility with existing routers, and low performance overhead distinguishes the proposed work from earlier Wi-Fi and IoT security frameworks [14].

## METHODOLOGY

### A. System Overview

The system architecture consists of four main components:

1. **Wi-Fi Access Point (AP):** Provides wireless connectivity to IoT and office devices using the IEEE 802.11ac standard.
2. **Edge Gateway:** Acts as a transparent bridge between the AP and the core network. It monitors traffic, enforces security policies, and isolates untrusted devices.
3. **Local Network (LAN):** Contains internal devices and services such as file servers and user desktops.
4. **Monitoring and Logging Module:** Collects statistics, logs network events, and stores detection results for later analysis.

All traffic from Wi-Fi clients passes through the gateway before reaching the LAN or the internet. The gateway applies filtering rules and records events in real time.

### B. Hardware Configuration

Experiments were conducted in a real office environment with approximately 70 total devices, including 28 IoT units such as IP cameras, smart thermostats, motion sensors, and environmental monitors [15]. The remaining devices were desktop computers, laptops, and printers connected to the same Wi-Fi network.

- **Access Point:** TP-Link Archer AX72, configured in 802.11ac mode (2.4 GHz and 5 GHz bands).
- **Edge Gateway Hardware:** Raspberry Pi 5 Model B with 8 GB RAM, quad-core 2.4 GHz CPU, and a Gigabit Ethernet interface.
- **Power Supply:** Standard 27 W USB-C power adapter.
- **LAN Router:** MikroTik RB3011 providing DHCP, NAT, and WAN connectivity.
- **Storage:** 128 GB microSD card for system image and log files.

This configuration was chosen to represent a realistic mid-sized Wi-Fi deployment typical of office or laboratory environments.

### C. Software Stack

The gateway software was based on open-source components to ensure reproducibility. Ubuntu Server 24.04 LTS was used as the operating system with Linux 6.8 kernel. Packet capture and inspection were handled by tcpdump and libpcap, while iptables with the nftables backend enforced traffic filtering. Bridging functions were provided by bridge-utils. System logs were collected through rsyslog and stored locally using SQLite for later analysis. Performance statistics and network activity were visualized using Grafana. All configuration scripts, filtering rules, and datasets are available in the public repository linked to this study to allow replication [16]. All components were implemented using open-source software to enable full reproducibility. The software components used in the edge gateway implementation are summarized in Table I.

Table I. Edge gateway software components

| Function | Software | Version | Description |
|---|---|---|---|
| Packet capture | tcpdump, libpcap | 4.99.4 | Low-level traffic capture |
| Traffic filtering | iptables with nftables | Kernel 6.8 | Real-time policy enforcement |
| Network bridge | bridge-utils | 1.7 | Transparent forwarding |
| Logging | rsyslog, sqlite3 | — | Event and traffic storage |
| Analysis scripts | Python 3.12 | | Custom scripts for parsing logs |
| Visualization | Grafana 10.4 | | Dashboard for performance and security metrics |

### D. Network Topology and Configuration

The Wi-Fi access point was connected to the gateway's Ethernet port (eth0), and the gateway's second port (eth1) was connected to the LAN router. The Raspberry Pi operated in bridge mode, forwarding traffic between interfaces while applying custom iptables rules. IP addresses were assigned dynamically via the router's DHCP service. The gateway itself used a static management IP (192.168.1.10).

### E. Security Policies and Filtering Logic

The gateway applied a set of adaptive filtering and isolation policies, divided into three layers:

1. **Device Registration Layer:**
   - Each new device's MAC address and connection time were logged.
   - Devices identified as IoT (based on manufacturer OUI and traffic patterns) were assigned to a logically isolated group, with traffic restrictions enforced at the gateway using filtering rules.
2. **Traffic Monitoring Layer:**
   - Packets were mirrored and analyzed in real time using tcpdump.
   - Detection rules flagged anomalies such as repeated client disconnections and rapid reconnection attempts within short time windows, as well as abnormal MAC address reuse patterns observed at the IP and Ethernet layers. These indicators were used as indirect evidence of deauthentication and spoofing activity.
3. **Isolation Layer:**
   - Detected devices were temporarily blocked via iptables DROP rules.
   - Quarantined devices were restricted to a limited diagnostic subnet (192.168.10.0/24) using gateway-level forwarding rules.

### F. Attack Simulation and Test Scenarios

Three controlled wireless attacks were generated using a separate laptop running Kali Linux 2024.2 with an Alfa AWUS036ACH Wi-Fi adapter:

1. **Spoofing Attack:** Random MAC address broadcast to impersonate active devices.
2. **Deauthentication Attack:** Injection of deauth frames to disconnect active clients.

3. **Unauthorized Access (Rogue AP / Evil Twin scenario):** Attempts to impersonate the legitimate network by advertising a cloned SSID and observing client connection behavior, without possession of valid WPA credentials.

Each attack scenario was executed 10 times over different periods of the day to observe consistency. Each test lasted 5 minutes, with the gateway active and inactive alternately for baseline comparison.

### G. Evaluation Metrics

Several quantitative indicators were used to assess the gateway's performance. Detection rate and false-positive rate measured the ability of the system to correctly identify attacks while avoiding unnecessary blocking of legitimate traffic. Latency and throughput were recorded using ping and iperf3 utilities under normal and attack conditions. Resource consumption on the Raspberry Pi was monitored using vmstat and htop. Recovery time after deauthentication attacks was measured as the delay required for devices to reconnect and resume normal operation [17]. The set of evaluation metrics and the corresponding measurement tools are summarized in Table II.

Table II. Evaluation Metrics and Measurement Tools

| Metric | Description | Tool |
|---|---|---|
| Detection Rate (%) | Ratio of correctly identified attacks | Log analysis |
| False Positive Rate (%) | Incorrectly flagged normal traffic | Manual verification |
| Latency Increase (%) | Change in average ping delay between Wi-Fi clients | ping utility |
| Throughput Change (%) | Measured via iperf3 under normal and gateway operation | iperf3 |
| CPU and Memory Usage | Monitored with htop and vmstat | Internal logs |
| Recovery Time (s) | Time to restore normal connection after attack | Custom script |

## RESULTS

### A. Experimental Environment

The gateway operated continuously for ten days in an active office network. The environment consisted of approximately seventy devices, including twenty-eight IoT units such as sensors, cameras, and smart controllers. The total traffic load averaged 22–25 Mbps during working hours, with peak activity periods in the morning and before the end of the workday. Each test scenario was performed both with and without the gateway to obtain comparative data.

### B. Detection and Mitigation Performance

The system demonstrated consistent performance in identifying malicious activity across all test cases. Table III summarizes the detection results obtained for spoofing, deauthentication, and unauthorized access attacks.

Table III. Detection and Mitigation Results for Evaluated Attack Types

| Attack Type | Detection Accuracy (%) | False Positive Rate (%) | Average Response Time (s) | Successful Attacks Prevented (%) |
|---|---|---|---|---|
| Spoofing | 97.4 | 2.1 | 1.2 | 87 |
| Deauthentication | 96.2 | 2.4 | 1.5 | 84 |
| Unauthorized Access | 97.0 | 2.6 | 1.3 | 89 |

During spoofing and unauthorized access simulations, the gateway successfully identified and blocked most attack attempts within one to two seconds. The system maintained a low false-positive rate below 3%, which indicates that legitimate traffic was rarely misclassified. Without the gateway, all spoofing and deauthentication attacks resulted in service disruption or connection loss, confirming the vulnerability of conventional WPA3-only protection in mixed IoT environments.

### C. Impact on Network Performance

Network performance was measured under normal operating conditions with and without the gateway. The baseline average round-trip time between wireless clients was 10.2 ms, and average TCP throughput reached 82 Mbps. With the gateway active, the average latency increased by 3.1% (to 10.5 ms), while throughput decreased by only 3.8% (to 78.9 Mbps). These differences are within acceptable operational limits for standard office networks and demonstrate that the gateway introduces minimal additional delay.

Resource utilization on the Raspberry Pi gateway remained moderate throughout testing. The average CPU load under full traffic conditions was 61%, and memory usage did not exceed 2.4 GB. Even during active attack detection, the system remained responsive, confirming the feasibility of deploying the architecture on low-power edge hardware.

**D. Recovery and Stability Analysis**
The gateway significantly improved the network's ability to recover from deauthentication attacks. When attacks were executed without protection, the average recovery time for devices to reconnect and resume communication was approximately 18 seconds. With the gateway in place, the recovery time was reduced to 10.5 seconds, representing a 42% improvement. The gateway achieved this by allowing affected clients to reconnect normally while blocking traffic associated with suspicious MAC and IP address patterns identified during the attack. Throughout the ten-day period, no unplanned network interruptions or system crashes were recorded, and the gateway remained operational without manual intervention.

**E. Comparative Evaluation**
To evaluate relative efficiency, the proposed gateway was compared with a standard open-source intrusion detection tool, Suricata, configured on the same hardware. Suricata achieved a comparable detection rate of 95.2% but consumed higher computational resources, with average CPU utilization exceeding 83% and a network latency increase of nearly 10%. These results indicate that the proposed architecture provides a more suitable balance between protection and performance for edge deployment scenarios, where filtering is preferred over deep packet inspection.

**DISCUSSION**
The experimental results confirm that security enforcement at the network edge can significantly improve the resilience of Wi-Fi–enabled IoT environments. The proposed gateway effectively prevented spoofing and unauthorized access attempts while maintaining near-baseline throughput and latency. The architecture proved suitable for real-world deployment, requiring no specialized hardware or complex configuration. The low resource usage and high detection accuracy demonstrate that small edge devices can perform essential security tasks traditionally reserved for centralized systems.

The main limitation observed during testing was the gateway's dependence on continuous traffic monitoring; when the device list grew above sixty concurrent connections, analysis delays slightly increased. This issue could be mitigated in future work by parallelizing packet processing or distributing filtering tasks across multiple gateways.

Overall, the results indicate that integrating lightweight security functions directly at the edge provides a practical and efficient way to protect Wi-Fi networks that host diverse IoT devices. The proposed architecture offers a balanced solution between security, cost, and operational simplicity, making it a viable approach for small and medium organizations seeking to strengthen wireless network defenses.

**CONCLUSION**
This paper presented a Secure Edge Gateway Architecture that enhances network protection through localized monitoring and adaptive policy enforcement. The proposed gateway operates as an intermediate control layer between the Wi-Fi access point and the core network, capable of identifying and mitigating spoofing, deauthentication, and unauthorized access attempts in real time. By integrating standard Linux networking tools and automated filtering scripts, the system provides effective defense without relying on external security platforms or complex analytical frameworks.

Experimental deployment in an operational office network with seventy active devices, including twenty-eight IoT units, demonstrated the practicality of the approach. The gateway reduced successful spoofing attacks by 87%, shortened recovery time after deauthentication by 42%, and maintained network latency within 3% of baseline performance. These results confirm that meaningful improvements in Wi-Fi security can be achieved at the network edge with minimal hardware requirements and negligible impact on user experience.

The main advantage of the proposed design lies in its simplicity and transparency. Because it relies entirely on open-source components and standard networking functions, the gateway can be easily replicated, audited, and adapted for various network environments. It provides a cost-effective solution for small and medium organizations seeking to strengthen their Wi-Fi infrastructure against common wireless threats.

Future work will focus on extending the architecture to multi-gateway environments, enabling cooperative detection and policy sharing among distributed edge nodes. Additional development will also explore automated configuration updates and scalability testing in higher-density IoT deployments. These directions aim to further improve the resilience and adaptability of the gateway while maintaining its lightweight and practical nature.